\RequirePackage{fix-cm}
\documentclass[smallextended]{svjour3}       % onecolumn (second format)
\smartqed  % flush right qed marks, e.g. at end of proof
\usepackage[T1]{fontenc}
\usepackage[latin9]{inputenc}
\usepackage{graphicx}
\usepackage{appendix}
\usepackage{tikz}
\usepackage{amsmath}
\usepackage{amssymb}
\usepackage{bm}
\usepackage[unicode=true,pdfusetitle,
 bookmarks=true,bookmarksnumbered=false,bookmarksopen=false,
 breaklinks=false,pdfborder={0 0 0},pdfborderstyle={},backref=false,colorlinks=true]
 {hyperref}
\usepackage{tabularx}
\usepackage{makecell}

\allowdisplaybreaks[4]

\begin{document}

\title{ Ancilla-assisted frequency estimation under phase covariant noises with Greenberger-Horne-Zeilinger states}
%\subtitle{Do you have a subtitle?\\ If so, write it here}

\author{Rui-Jie Cai  \and  Wei Zhong \and Lan Zhou  \and Yu-Bo Sheng}

%\authorrunning{Short form of author list} % if too long for running head

\institute{ R. J. Cai  \at
Institute of Quantum Information and Technology, Nanjing University
of Posts and Telecommunications, Nanjing 210003, China\\
\and
 W. Zhong \at
Institute of Quantum Information and Technology, Nanjing University
of Posts and Telecommunications, Nanjing 210003, China\\
National Laboratory of Solid State Microstructures, Nanjing University,
Nanjing 210093, China\\
 \email{zhongwei1118@gmail.com}
 \and
 L. Zhou \at
School of Science, Nanjing University of Posts and Telecommunications,
Nanjing 210003, China\\
\and
 Y. B. Sheng \at
Institute of Quantum Information and Technology, Nanjing University
of Posts and Telecommunications, Nanjing 210003, China\\
}

\date{Received: date / Accepted: date}
% The correct dates will be entered by the editor

\maketitle

\begin{abstract}
It has been demonstrated that the optimal sensitivity achievable with
Greenberger-Horne-Zeilinger states is the same as that with uncorrelated
probes in the frequency estimation in the presence of uncorrelated
Markovian dephasing {[}S. F. Huelga, \emph{et al.}, Phys. Rev. Lett.
\textbf{79}, 3865 (1997){]}. Here, we extend this issue by examining
the optimal frequency sensitivities achievable by the use of ancilla-assisted
strategy, which has been proposed recently for robust phase estimation.
We present the ultimate frequency sensitivities bounded by the quantum
Fisher information for a general case in the presence of Markovian
covariant phase noises, and the optimal measurement observables that
can saturate the theoretical sensitivity bounds. We also demonstrate
the effectiveness of the ancilla-assisted strategy for preserving
frequency sensitivities suffering from specific physically ground
noises.
\end{abstract}

%\keywords:{Frequency estimation, \and Markovian covariant noises, \and Greenberger-Horne-Zeilinger states}
%\begin{keyword}
%Greenber-Horne-Zeilinger state, uncorrelated probes, Markovian dephasing
%03.67.Pp, 03.67.Hk, 03.65.Ud
%\end{keyword}
\maketitle

\section{Introduction}
Quantum enhanced metrology plays an important role in super-precision
measurement applications, such as atomic frequency standards \cite{Wineland1994PRA,Bollinger1996PRA,Leibfried2004SCI},
quantum magnetometers \cite{Taylor2008NP,Goldstein2011PRL}, gravitational
wave detection \cite{Collaboration2011NP,Aasi2013NP,Barsotti2018RP},
quantum gyroscopes \cite{Halkyard2010PRA,Helm2015PRL,Stevenson2015PRL}
etc. It has been attracted intense attention owing to its potential
ability to improve measurement precision by a factor of $\sqrt{N}$
improvement over the standard quantum limit $1/\sqrt{N}$, where $N$
is the number of samples, such as atoms or photons \cite{Giovannetti2006PRL,Giovannetti2011}.
This is well-known Heisenberg limit $1/N$ and attainable by means
of Greenberger-Horne-Zeilinger (GHZ) states \cite{Giovannetti2006PRL,Giovannetti2011}.
However, these exotic states are of extremely delicate and fragile.
They may be easily destroyed by inevitabl
e noises in realistic experiments
and lose its capacity in sensitivity enhancement \cite{Froewis2011PRL,Ma2011PRA,Chaves2012PRA,Zhong2013PRA}.
In 1997, Huelga \emph{et al.} first found that, in frequency estimation,
the sensitivity achievable with GHZ states is the same as that with
uncorrelated probes in the presence of the uncorrelated Markovian
dephasing noise \cite{Huelga1997PRL}. Recently, more generalized
results indicated that any small amount of realistic noise restricts
the advantage of quantum strategies to an improvement by a constant
factor \cite{Escher2011NP,Demkowicz-Dobrzanski2012nc}.

Under pressure by these no-go results, various strategies have been
put forward in the literature to mitigate the detrimental effects
of noise on sensitivities, including quantum error correction \cite{Duer2014PRL,Arrad2014PRL,Kessler2014PRL,Lu2015NC,Herrera-Marti2015PRL,Unden2016PRL,Layden2019PRL,Sekatski2017quantummetrology},
dynamical decoupling \cite{Sekatski2016NJP,Tan2013PRA}, time-continuous
monitoring \cite{Gammelmark2013PRA,Gammelmark2014PRL,Catana2014PRA,Kiilerich2014PRA,Kiilerich2016PRA,Plenio2016PRA,Cortez2017PRA,Chase2009PRA,Stockton2004PRA,Geremia2003PRL,Albarelli2017NJP}
etc. Among these, a novel ancilla-assisted strategy has been proposed
recently by entangling the probes with the ancillary qubits which
are free from phase encoding and decoherence \cite{Demkowicz-Dobrzafmmodecutenlseniski2014PRL,Huang2016PRA,Huang2018PRA}.
This strategy has been identified as an effective way of preserving
sensitivities in phase estimation under some noisy channels and also
experimentally demonstrated \cite{Sbroscia2018PRA,Wang2018PRA}.
These analyses were carried out only for the case of phase estimation,
but there is no evidence on the effectiveness of the ancilla-assisted
strategy for frequency estimation. More recently, some evidences have
shown that the ultimate lower sensitivity bounds in noisy frequency
estimation with and without ancillary qubits are the same in the asymptotic
limit of $N\rightarrow\infty$ \cite{Smirne2016PRL}. It is still
ambiguous whether the ancilla-assisted strategy is useful or not for
noisy frequency estimation with finite number of probes.

In this work, we address this issue by examining the sensitivities
achievable with generalized GHZ states in frequency estimation suffering
from a general class of Markovian noises, of which the processes commute
with the unitary evolution governed by the system Hamiltonian, which
is referred as phase covariant noises \cite{Smirne2016PRL}. To identify
the effectiveness of the ancilla-assisted strategy in noisy frequency
estimation, we preform our study based on quantum estimation theory
\cite{Helstrom1976Book,Holevo1982Book,Braunstein1994PRL} and exactly
obtain the ultimate frequency sensitivities achievable in aforementioned
noisy scenarios. Meanwhile, the measurement observables that approach
the sensitivity bounds to frequency are also presented. We then specialize
these results to the most popular three kinds of decoherence models,
such as phase damping channel (DPC), and amplitude-damping channel
(PDC) and depolarizing channel (DPC), and obtain the optimal frequency
sensitivities in these three cases. We also compare the performances
of two strategies equipped with and without ancillary qubits.

This paper is organized as follows. In Sec. II, we briefly introduce
background information about quantum frequency estimation in the presence
of phase covariant noises. We then derive the frequency sensitivities
attainable with generalized GHZ states for both ancilla-free and ancilla-assisted
strategies and show the optimal measurement observables to access
these sensitivities. In Sec. III, we specialize into three conventional
quantum channels, such as ADC, PDC, and DPC. Finally, our conclusions
are made in Sec. IV. In appendix, some derivations are presented in
detail.
\begin{figure}[b]
\begin{center}
\includegraphics[scale=0.8]{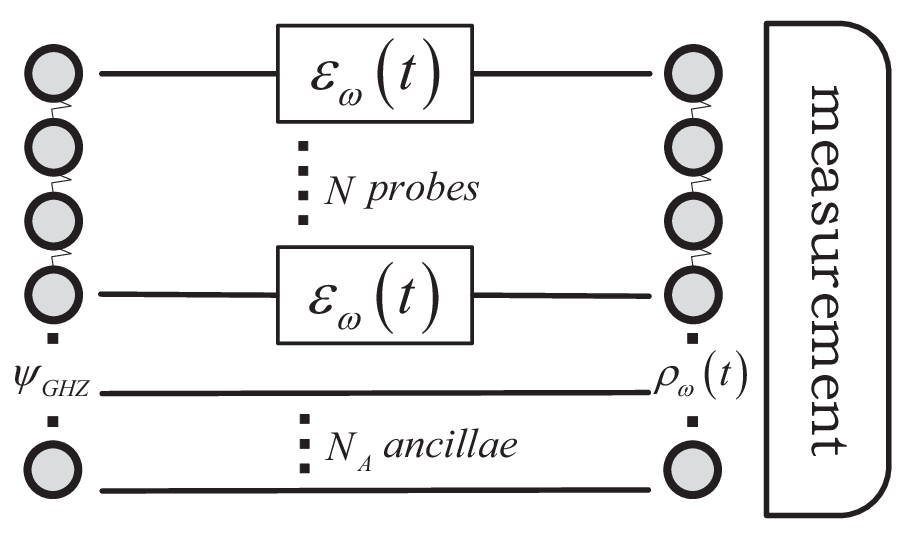}\caption{(Color online) Schematic of noisy frequency estimation with generalized
GHZ states. \label{fig:EAM}}
\end{center}
\end{figure}

%%ÕýÎÄ
\section{Noisy quantum frequency estimation}
\subsection{Frequency estimation in the presence of phase covariant noises}

In general, an unknown parameter $\omega$ (specifically the detuning
between the frequency of the atomic transition and the frequency of
the driving field) to be estimated is unitarily encoded on $N$ probes,
i.e., two-level atoms or qubits, of which the initial state is denoted
by $\rho\left(0\right)$, over the interrogation time $t$ \cite{Huelga1997PRL}.
After the encoding process, the value of $\omega$ is inferred from
the outcomes of the measurement on the final state of the system $\rho_{\omega}\left(t\right)$.
To accurately infer its value, these processes are required to be
repeated many times denoted as $\upsilon$, and the total experiment
time is taken as $T=\upsilon t$. Hence, the two quantities $N$ and
$T$ suggest the total resource consumed in a frequency estimation
task. According to quantum estimation theory, the theoretical sensitivity
limit to an unbiased frequency estimator $\hat{\omega}$ is bounded
by the inverse of quantum Fisher information (QFI) \cite{Helstrom1976Book,Holevo1982Book,Braunstein1994PRL}
\begin{equation}
\Delta^{2}\hat{\omega}T\geq\frac{t}{F\left[\rho_{\omega}\left(t\right)\right]}.\label{eq:QCRB}
\end{equation}
It means that the larger value of the QFI is the higher sensitivity
could acquire. The above bound could be attained by the use of the
maximal likelihood estimator for sufficiently large number of repetitions
of the experiment \cite{Helstrom1976Book,Holevo1982Book,Braunstein1994PRL}.

In realistic settings, it is inevitable that noises will be introduced
to the encoding process and negatively affect the estimation sensitivity
that one can obtain. Suppose that the evolution of each probe is independent
identical. Since any quantum process can be described as a completely
positive and trace-preserving (CPT) map $\mathcal{E}$, thus a noisy
encoding process can be expressed by
\begin{equation}
\rho_{\omega}\left(t\right)=\mathcal{E}_{\omega}\left(t\right)^{\otimes N}\rho\left(0\right),\label{eq:noisy-encoding-map}
\end{equation}
where $\rho\left(0\right)$ denotes the initial state of the system
and $\mathcal{E}_{\omega}\left(t\right)$ represents the dynamical
map on a single qubit. Here, we concentrate on a general class of
phase covariant maps requiring that a CPT map $\mathcal{E}_{\omega}\left(t\right)$
can be divided into two parts: unitary encoding $\mathcal{U}_{\omega}\left(t\right)$
and frequency-independent dissipative part $\Gamma\left(t\right)$
and they commute with each other, i.e.,
\begin{equation}
\mathcal{E}_{\omega}\left(t\right)=\Gamma\left(t\right)\circ\mathcal{U}_{\omega}\left(t\right)=\mathcal{U}_{\omega}\circ\Gamma\left(t\right).
\end{equation}
This means that one can freely exchange the order of the operations
between the encoding and the dissipation.

To intuitively describe the noisy encoding processes, we take a geometric
point of view in the way a single qubit state is denoted by a real
vector $\bm{r}$ in the space of Pauli matrices, which is known as
the so-called Bloch vector. Under the Bloch representation, the phase
covariant map $\mathcal{E}_{\omega}\left(t\right)$ can be equivalently
represented by an affine map \cite{Smirne2016PRL,Zhong2013PRA}
\begin{equation}
\mathcal{E}_{\omega}\left(t\right):\bm{r}\mapsto A\bm{r}+\bm{c},\label{eq:affine}
\end{equation}
where the transformation matrix $A$ is a $3\times3$ real matrix
depicted by
\begin{equation}
A=\left(\begin{array}{ccc}
\eta_{\perp}\left(t\right)\cos\theta\left(t\right) & -\eta_{\perp}\left(t\right)\sin\theta\left(t\right) & 0\\
\eta_{\perp}\left(t\right)\sin\theta\left(t\right) & \eta_{\perp}\left(t\right)\cos\theta\left(t\right) & 0\\
0 & 0 & \eta_{\parallel}\left(t\right)
\end{array}\right),\label{eq:affineA}
\end{equation}
and the translation vector $\bm{c}$ is given as
\begin{equation}
\bm{c}=\left(\begin{array}{ccc}
0, & 0, & \kappa\left(t\right)\end{array}\right)^{{\rm T}}.\label{eq:affinec}
\end{equation}
The transformation matrix $A$ refers to a rotation of the Bloch sphere
around the $z$ axis by a angle of $\theta\left(t\right)=\phi\left(t\right)+\omega t$,
a shrinking of the Bloch sphere in $xy$ plane by a factor $0\leq\eta_{\perp}\left(t\right)\leq1,$
and a shrinking of the $x$ axis by a factor $0\leq\eta_{\parallel}\left(t\right)\leq1$
\cite{Smirne2016PRL}. The translation vector $\bm{c}$ refers to
a displacement in the $z$ direction by a factor of $0\leq\kappa\left(t\right)\leq1$
\cite{Smirne2016PRL}. This kind of phase covariant map encompasses
a broad class of Markovian noises taking place in physical ground,
such as ADC, DPC and PDC, as will discussed in Sec.~III.

\subsection{Frequency sensitivities for ancilla-free strategy}
Now we discuss the obtainable frequency sensitivities in the above
noisy scenarios when the initial state of the system of $N$ qubits
is assumed to be a generalized GHZ state $\rho_{{\rm GHZ}}=\vert\psi_{{\rm GHZ}}\rangle\langle\psi_{{\rm GHZ}}\vert$
with
\begin{equation}
\vert\psi_{{\rm GHZ}}\rangle=c_{1}\vert0\rangle^{\otimes N}+c_{2}\vert1\rangle^{\otimes N},\label{eq:GHZ}
\end{equation}
satisfying the normalization condition of $\left|c_{1}\right|^{2}+\left|c_{2}\right|^{2}=1$.
It corresponds to the maximally entangled states when $c_{1}=c_{2}=1/\sqrt{2}.$
Under the action of noisy phase encoding of Eq.~\eqref{eq:noisy-encoding-map},
the evolved state from $\rho_{{\rm GHZ}}$ can be expressed as the
direct sum form of $\rho_{\omega}\left(t\right)=\varrho_{\omega}\oplus\varrho_{D}$
\cite{Ma2011PRA,Smirne2016PRL}, where $\varrho_{\omega}$ and $\varrho_{D}$
are $2$-dimensional matrix and $\left(N^{2}-2\right)$-dimension
diagonal matrix, respectively, as following
\begin{eqnarray}
\varrho_{\omega} & = & \left(\begin{array}{cc}
2^{-N}\left(\left|c_{1}\right|^{2}A_{++}^{N}+\left|c_{2}\right|^{2}A_{-+}^{N}\right) & c_{1}c_{2}^{\ast}\eta_{\perp}^{N}e^{-iN\left[\theta\left(t\right)+\omega t\right]}\\
c_{1}^{\ast}c_{2}\eta_{\perp}^{N}e^{iN\left[\theta\left(t\right)+\omega t\right]} & 2^{-N}\left(\left|c_{1}\right|^{2}A_{--}^{N}+\left|c_{2}\right|^{2}A_{+-}^{N}\right)
\end{array}\right),\label{eq:rho-2-free}\\
\varrho_{D} & = & 2^{-N}\sum_{k=1}^{N-1}\binom{N}{k}\left(\vert c_{1}\vert^{2}A_{++}^{k}A_{--}^{N-k}+\left|c_{2}\right|^{2}A_{-+}^{k}A_{+-}^{N-k}\right)\vert0\rangle\langle0\vert^{\otimes k}\vert1\rangle\langle1\vert^{\otimes N-k},\quad\quad\label{eq:rho-d-free}
\end{eqnarray}
by setting $A_{\pm\pm}=1\pm\eta_{\parallel}\pm\kappa$. Due to being
independent of $\omega$, the sub-matrix $\varrho_{D}$ makes no contribution
to the QFI and hence is irrelevant to the frequency sensitivity of
Eq.~\eqref{eq:QCRB}. As detailed shown in appendix A, we present
the expression of the QFI for the generalized GHZ state under covariant
noises as
\begin{eqnarray}
F_{N} & = & \frac{4t^{2}\left|c_{1}\right|^{2}\left|c_{2}\right|^{2}N^{2}\eta_{\perp}^{2N}}{2^{-N}\left[\left|c_{1}\right|^{2}\left(A_{++}^{N}+A_{--}^{N}\right)+\left|c_{2}\right|^{2}\left(A_{-+}^{N}+A_{+-}^{N}\right)\right]},\label{eq:QFI-noisyGHZ}
\end{eqnarray}
This general expression reduces to the result shown in Ref.~\cite{Smirne2016PRL}
by setting $c_{1}=c_{2}=1/\sqrt{2}$. By replacing $N$ with $1$
in Eq.~\eqref{eq:QFI-noisyGHZ} and then multiplying by $N$ as a
result of the additivity of the QFI that $F\left(\rho\right)=\sum_{i}F\left(\varrho_{i}\right)$
for product states $\rho=\sum_{i}\varrho_{i}$, one can directly obtain
the QFI for uncorrelated probe states as
\begin{eqnarray}
F_{U} & = & \frac{4t^{2}\left|c_{1}\right|^{2}\left|c_{2}\right|^{2}N\eta_{\perp}^{2}}{2^{-1}\left[\left|c_{1}\right|^{2}\left(A_{++}^{N}+A_{--}^{N}\right)+\left|c_{2}\right|^{2}\left(A_{-+}^{N}+A_{+-}^{N}\right)\right]}.\label{eq:QFI-noisyuncorrelated}
\end{eqnarray}

It is well known that, in the case of absence of noises, GHZ states
permit a Heisenberg-scaling sensitivity to frequency. This sensitivity
can be attained by the use of parity measurement scheme. However,
in the noisy situations as discussed above, parity measurement fails
to access the ultimate sensitivities given by Eqs.~\eqref{eq:QCRB}
and \eqref{eq:QFI-noisyGHZ}. Here, we find a measurement observable
\begin{equation}
O_{N}=\sigma_{x}\oplus\bm{1}_{N^{2}-2}=\vert0\rangle\langle1\vert^{\otimes N}+\vert1\rangle\langle0\vert^{\otimes N},\label{eq:parity}
\end{equation}
which is always optimal in our cases under the condition of $\theta\left(t\right)+\omega t=2k\pi/N,k\in\mathbb{Z}$
(see appendix B for details). This measurement scheme can be understood
in principle by a simultaneous measurement of the two matrix elements
of $\rho_{\omega}\left(t\right)$ placed at $\vert0\rangle\langle1\vert^{\otimes N}$
and $\vert1\rangle\langle0\vert^{\otimes N}$ with quantum tomographic
technique \cite{James2001PRA}.

\subsection{Frequency sensitivities for ancilla-assisted strategy}

Below, we consider an ancilla-assisted strategy that allows $N_{A}$
ancillary qubits to be initially correlated with the probes and detected
at the end of protocol, but free from the frequency encoding and noises,
see Fig.~\ref{fig:EAM}. The resulting state of the composite system
between probes and ancillary qubits can be described by
\begin{equation}
\rho_{\omega}^{A}\left(t\right)=\mathcal{E}_{\omega}\left(t\right)^{\otimes N}\otimes\bm{1}^{\otimes N_{A}}\left[\rho_{{\rm SA}}\left(0\right)\right],
\end{equation}
where the initial state $\rho_{{\rm SA}}\left(0\right)$ is still
assumed to be prepared in the generalized GHZ states as Eq.~\eqref{eq:GHZ}
by replacing $N$ with $N+N_{A}$. Similar to the ancilla-free case
discussed previously, the evolved state $\rho_{\omega}^{A}\left(t\right)$
can still be written in the direct sum form of $\rho_{\omega}^{A}\left(t\right)=\varrho_{\omega}^{A}\oplus\varrho_{D}^{A}$
with
\begin{eqnarray}
\varrho_{\omega}^{A} & = & \left(\begin{array}{cc}
2^{-N}\left|c_{1}\right|^{2}A_{++}^{N} & c_{1}c_{2}^{\ast}\eta_{\perp}^{N}e^{-iN\left[\theta\left(t\right)+\omega t\right]}\\
c_{1}^{\ast}c_{2}\eta_{\perp}^{N}e^{iN\left[\theta\left(t\right)+\omega t\right]} & 2^{-N}\left|c_{2}\right|^{2}A_{+-}^{N}
\end{array}\right),\label{eq:rho-2-ancilla}\\
\varrho_{D}^{A} & = & 2^{-N}\bigg\{\sum_{k=1}^{N}\binom{N}{k}\left|c_{1}\right|^{2}A_{++}^{k}A_{--}^{N-k}\vert0\rangle\langle0\vert^{\otimes N_{A}+k}\vert1\rangle\langle1\vert^{\otimes N-k} \nonumber\\
&&+\sum_{k=0}^{N-1}\binom{N}{k}\left|c_{2}\right|^{2}A_{-+}^{k}A_{+-}^{N-k}\vert0\rangle\langle0\vert^{\otimes k}\vert1\rangle\langle1\vert^{\otimes N_{A}+N-k}\bigg\}.
\end{eqnarray}

Following the same procedure used in deriving Eq.~\eqref{eq:QFI-noisyGHZ}
(see appendix A for detailed derivation), we obtain the QFI about
$\rho_{\omega}^{A}\left(t\right)$ as
\begin{eqnarray}
F_{A} & = & \frac{4t^{2}\left|c_{1}\right|^{2}\left|c_{2}\right|^{2}N^{2}\eta_{\perp}^{2N}}{2^{-N}\left(\left|c_{1}\right|^{2}A_{++}^{N}+\left|c_{2}\right|^{2}A_{+-}^{N}\right)}.\label{eq:QFI-ansillary-noisyGHZ}
\end{eqnarray}
This is a key result of the paper. The above expression subtly differs
from Eq.~\eqref{eq:QFI-noisyGHZ} by absence of two terms relevant
to $A_{--}^{N}$ and $A_{-+}^{N}$ in the denominator. By a brief
glance at the expression of Eq.~\eqref{eq:QFI-ansillary-noisyGHZ},
one could find that it is independent on the number of ancillary qubits.
It means that the ancilla-assisted strategy with single ancillary
qubit performs the same behavior as the one by using more ancillary
qubits in noisy frequency estimation. In other words, there is no
any benefit to the sensitivity enhancement through increasing the
number of ancillary qubits. Similar things also happens in noisy phase
estimation. Thus it is not necessary to use an equal number of ancillae
and probes in noisy phase estimation when the GHZ states are taken
as the initial state, as experimentally implemented in Ref.~\cite{Wang2018PRA}.
Moreover, the measurement observable similar to Eq.~\eqref{eq:parity}
as
\begin{equation}
O_{A}=\sigma_{x}\oplus\bm{1}_{\left(N+N_{A}\right)^{2}-2}=\vert0\rangle\langle1\vert^{\otimes N+N_{A}}+\vert1\rangle\langle0\vert^{\otimes N+N_{A}},\label{eq:Aparity}
\end{equation}
can still saturate the sensitivity limits given by Eq.~\eqref{eq:QFI-ansillary-noisyGHZ}
under the same condition of $\theta\left(t\right)+\omega t=2k\pi/N,k\in\mathbb{Z}$
as for ancilla-free case.

\section{Optimal frequency sensitivities under specific noises}

In this section, we discuss the sensitivities achievable with the
ancilla-assisted strategy under three types of noises, including,
ADC, DPC, and PDC, which are most popular in physical ground. They
 are regularly depicted by the following master equations with constant
decay rate, respectively \cite{Romero2012PS},
\begin{eqnarray}
\frac{d\rho\left(t\right)}{dt} & = & -\frac{i}{2}\left[\omega\sigma_{z},\rho\left(t\right)\right]+\gamma\left(\sigma_{-}\rho\left(t\right)\sigma_{+}-\frac{1}{2}\left\{ \sigma_{+}\sigma_{-},\rho\left(t\right)\right\} \right),\\
\frac{d\rho\left(t\right)}{dt} & = & -\frac{i}{2}\left[\omega\sigma_{z},\rho\left(t\right)\right]+\frac{1}{4}\gamma\sum_{i=x,y,z}\left(\sigma_{i}\rho\left(t\right)\sigma_{i}-\rho\left(t\right)\right),\\
\frac{d\rho\left(t\right)}{dt} & = & -\frac{i}{2}\left[\omega\sigma_{z},\rho\left(t\right)\right]+\frac{1}{2}\gamma\left(\sigma_{z}\rho\left(t\right)\sigma_{z}-\rho\left(t\right)\right).
\end{eqnarray}
According to the relation between the master equation and affine map
given in \cite{Smirne2016PRL}, one can determine the variable coefficients
$\theta,\eta_{\perp},\eta_{\parallel},$ and $\kappa$ presented in
Eq.~\eqref{eq:affineA} and \eqref{eq:affinec} corresponding to
the above three noise models as: (1) $\theta=0$, $\eta_{\perp}=\exp\left(-\gamma t/2\right)$,
$\eta_{\parallel}=\exp\left(-\gamma t\right)$, and $\kappa=\exp\left(-\gamma t\right)-1$
for ADC; (2) $\theta=\kappa=0$ and $\eta_{\perp}=\eta_{\parallel}=\exp\left(-\gamma t\right)$
for DPC; (3) $\theta=\kappa=0,$ and $\eta_{\perp}=\exp\left(-\gamma t\right)$,
and $\eta_{\parallel}=1$ for PDC. With Eqs.~\eqref{eq:QFI-noisyGHZ},
\eqref{eq:QFI-noisyuncorrelated}, and \eqref{eq:QFI-ansillary-noisyGHZ},
we obtain the QFIs about the uncorrelated state and the maximally
entangled state ($c_{1}=c_{2}=1/\sqrt{2}$) in the cases of in the
presence of ADC, DPC, and PDC, as shown in Table $1$.

\begin{table}[tb]
\caption{The QFIs for the maximally entangled probe states under three different
kinds of decoherence channels with and without ancillary qubits. \label{tab:measurements}}

\begin{tabular}{cccc}
\hline
\hline
\\[2pt]
Noise model & $F_{N}/t$ & $F_{A}/t$ & $F_{U}/t$\\
\\[2pt]
\hline
\\[2pt]
ADC& $\frac{2tN^{2}e^{-N\gamma t}}{1+e^{-N\gamma t}+\left(\left(1-e^{-\gamma t}\right)^{N}\right)}$ & $\frac{2tN^{2}e^{-N\gamma t}}{1+e^{-N\gamma t}}$ & $tNe^{-\gamma t}$ \\
\\[3pt]
DPC& $\frac{2tN^{2}e^{-2N\gamma t}}{\left(1-\frac{1-e^{-\gamma t}}{2}\right)^{N}+\left(\frac{1-e^{-\gamma t}}{2}\right)^{N}}$ & $\frac{tN^{2}e^{-2N\gamma t}}{\left(1-\frac{1-e^{-\gamma t}}{2}\right)^{N}}$ & $tNe^{-2\gamma t}$\\
\\[3pt]
PDC& $tN^{2}e^{-2N\gamma t}$ & $tN^{2}e^{-2N\gamma t}$ & $tNe^{-2\gamma t}$\\[2pt]
\hline
\hline
\end{tabular}
\end{table}

To clearly quantify the effect of ancilla-assisted strategy in noisy
frequency estimation, we introduce a ratio between the sensitivities
achievable with the entangled and uncorrelated probe states, defined
by
\begin{equation}
R=\frac{\min\left(\Delta^{2}\hat{\omega}\right)_{G}}{\min\left(\Delta^{2}\hat{\omega}\right)_{U}}=\frac{\max_{t}\left(F_{U}/t\right)}{\max_{t}\left(F_{G}/t\right)},\label{eq:ratio}
\end{equation}
where the subscripts $G$ and $U$ denote the GHZ- and uncorrelated-probe
protocols, respectively. According to the QCRB of Eq.~\eqref{eq:QCRB},
the frequency sensitivity is determined by the inverse of $F/t$.
In the presence of noises, the quantity of $F/t$ always functions
as $te^{-t\beta\left(\gamma\right)}$ up to a constant factor relevant
to $N$, as shown in Table $1$, and has a maximum point at certain
value of the interrogation time $t=t_{{\rm opt}}\left(\beta,N\right)$.
As for the uncorrelated protocol, for instance, $F_{U}/t$ gets the
maximum value $N/\left(e\gamma\right)$ at $t_{{\rm opt}}=1/\gamma$
for the case of ADC and $N/\left(2e\gamma\right)$ at $t_{{\rm opt}}=1/\left(2\gamma\right)$
for the cases of DPC and PDC. As for the entangled protocol, it is
more complicated to analytically maximizing $F/t$ over $t$ in the
ADC and DPC cases, except for PDC, in which case, the maximum value
of $F_{G}/t$ is identical to that for the uncorrelated one, thus
having $R=1$ \cite{Huelga1997PRL}, and is placed at a $N$-dependent
optimal interrogation time $t_{{\rm opt}}=1/\left(2N\gamma\right)$.

We plot the sensitivity raitos $R$ and the optimal interrogation
times $t_{{\rm opt}}$ for the ADC and DPC cases in Fig.~2. It is
depicted that the ratios $R$ by applying ancilla-free strategy decreases
from $1$ and rapidly reach the plateaus $R=0.66$ and $0.76$ for
ADC and DPC, respectively. Correspondingly, the optimal interrogation
time $t_{{\rm opt}}$ in both two cases decrease exponentially as
$N$ increases. While for the ancilla-assisted strategy the ratio
$R$ is invariant at $R=0.66$ for ADC and nearly invariant at $R=0.76$
for DPC, except at $N=1$ where $R=0.78$. In contrast to the ancilla-free
one, the enhancement in sensitivity by ancilla-assisted strategy the
is apparent considerably in few number of probes, e.g., $N=1$ and
$2$, but fade with a modest increasing of $N$. The behavior of $t_{{\rm opt}}$
for the ancilla-assisted strategy is similar to that for the ancilla-less
one, where $t_{{\rm opt}}$ in the former case is slightly longer
than that the latter case at few $N$ and they merge as $N$ increases.

\begin{figure}[tb]
\begin{centering}
\includegraphics[scale=0.22]{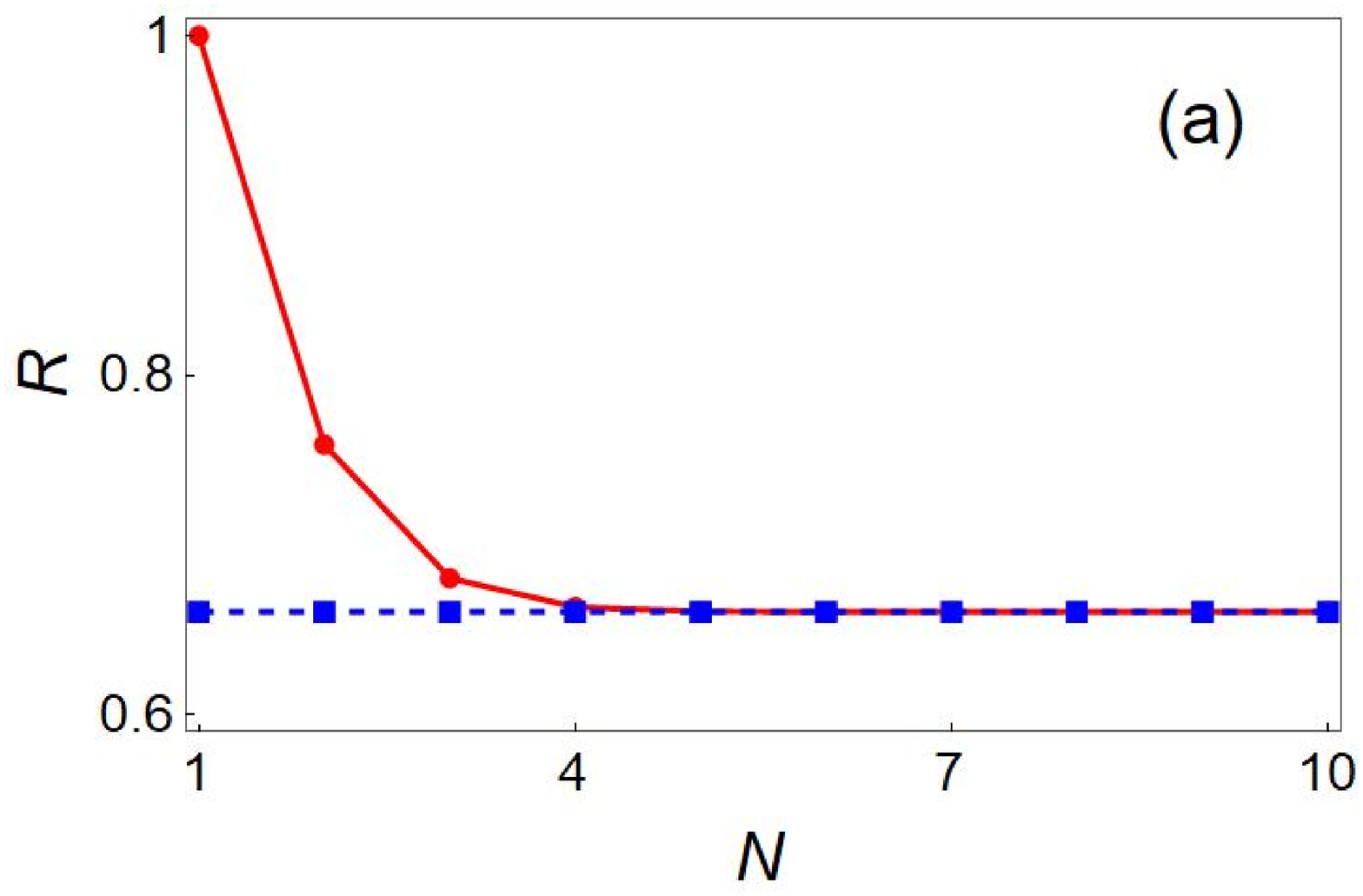}\,\includegraphics[scale=0.22]{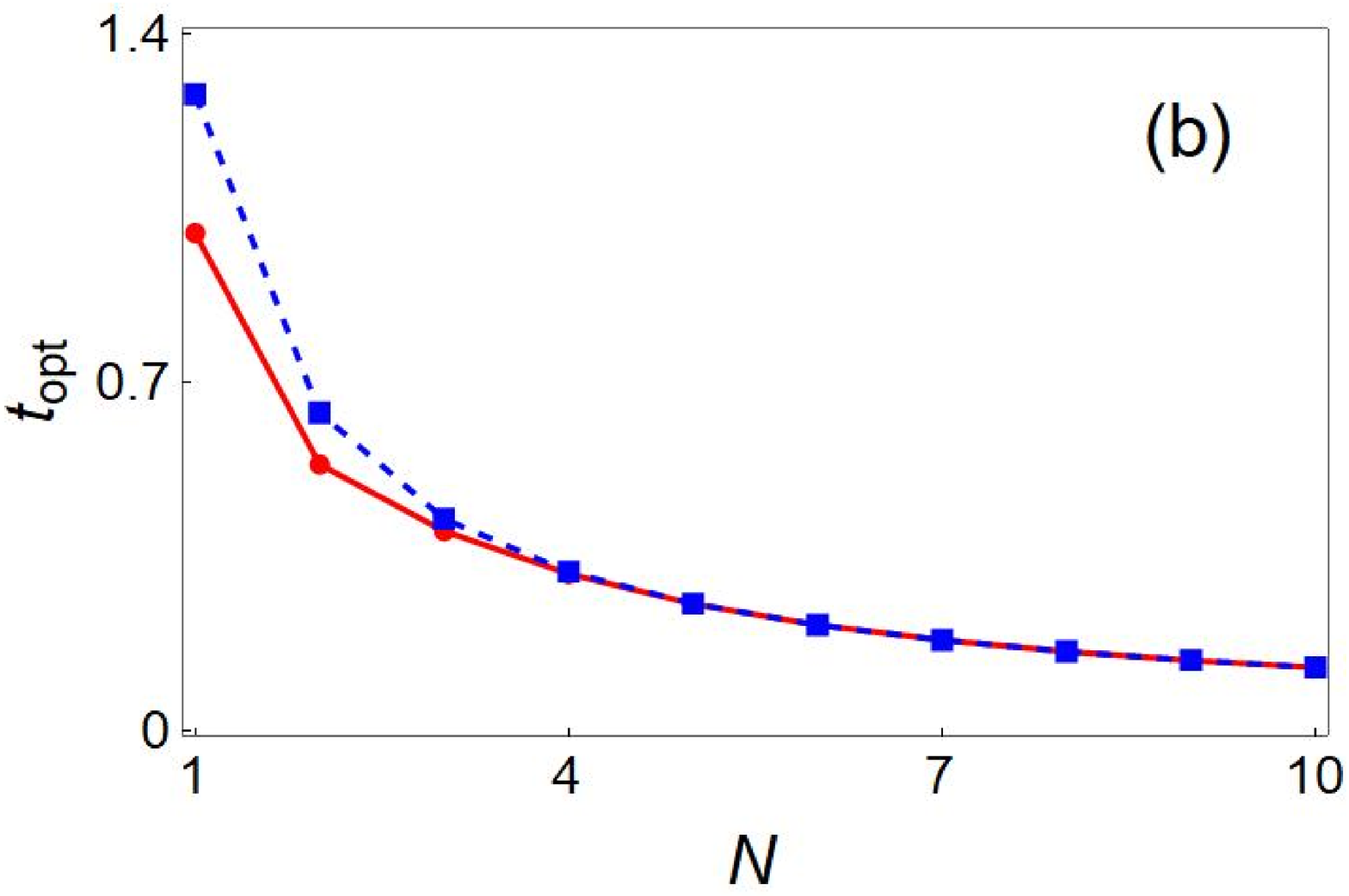}
\par\end{centering}
\centering{}\includegraphics[scale=0.22]{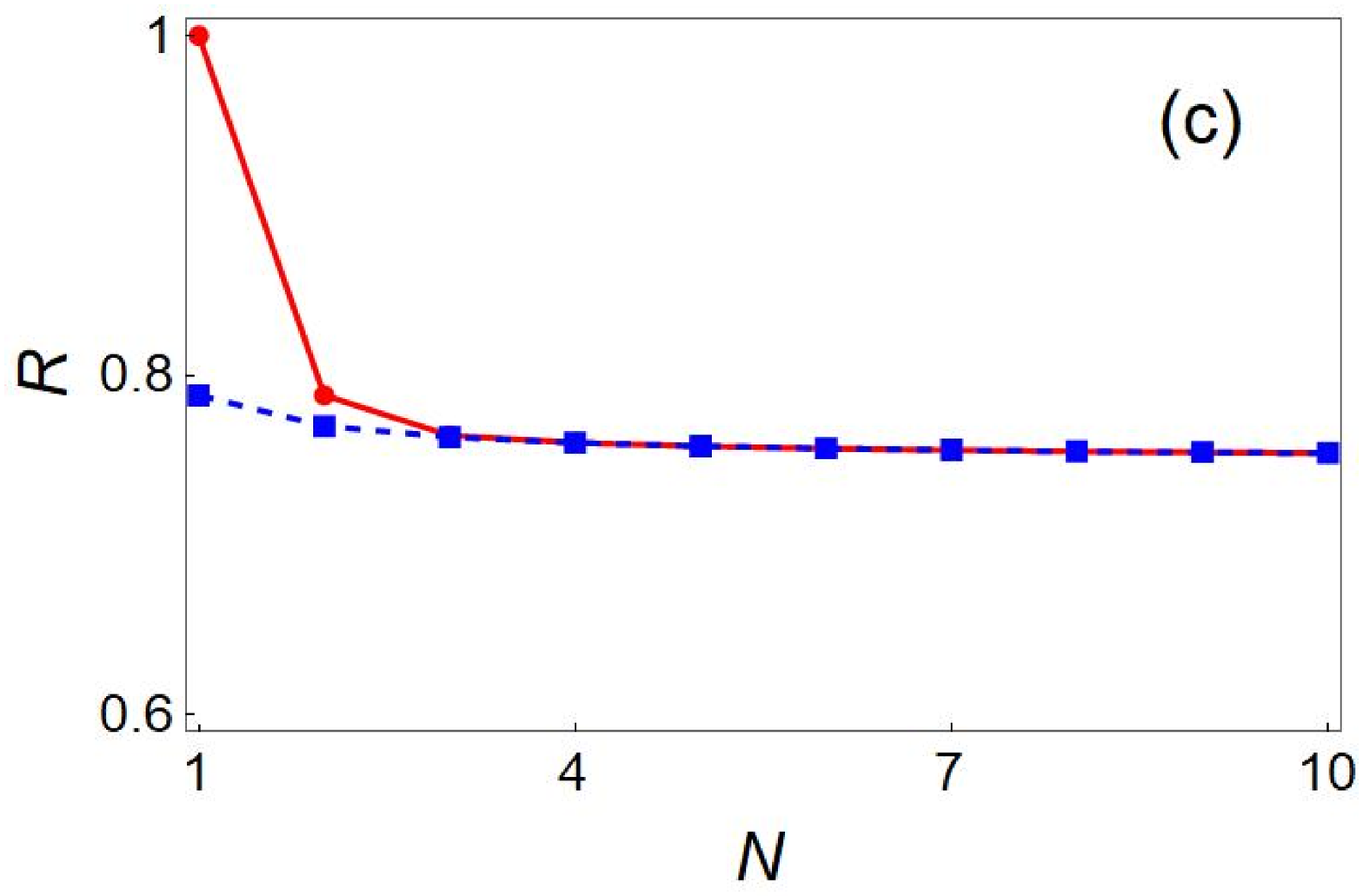}\,\includegraphics[scale=0.22]{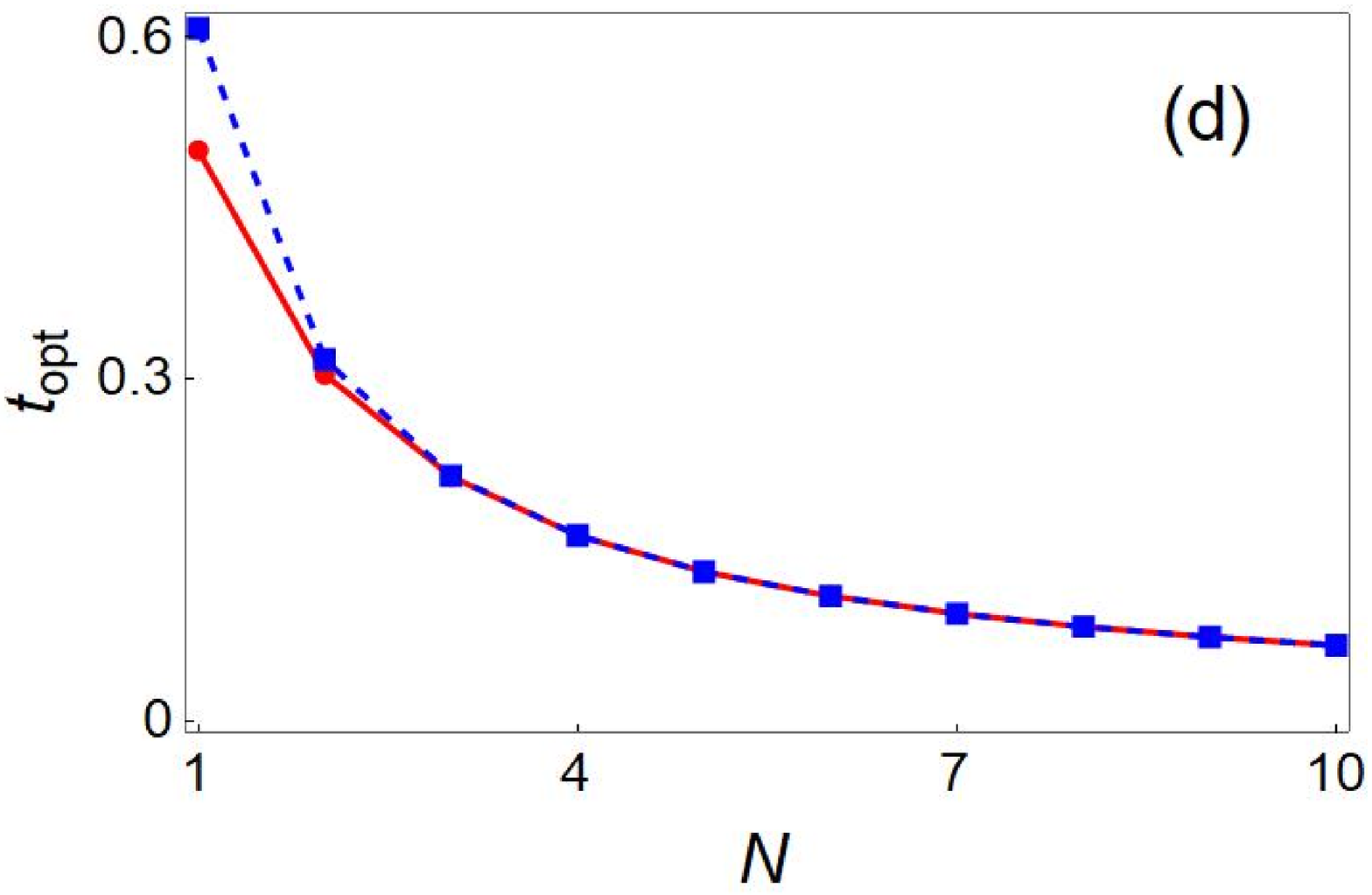}\caption{(Color online) The sensitivity ratios $R$ defined by Eq.~\eqref{eq:ratio}
and the optimal interrogation time $t_{{\rm opt}}$ as a function
of the probe number $N$ for the ADC (a, b) and DPC (c, d) cases.
Red solid lines correspond to the ancilla-free strategy and blue dashed
lines to the ancilla-assisted one. \label{fig:sensitivity}}
\end{figure}

\section{Conclusion}
We investigated the frequency sensitivity achievable with generalized
GHZ states in the presence of Markovian covariant phase noises, which
models a broad family of noises requiring that the processes between
the system Hamiltonian and dissipation commute. By invoking the QFI
theory, we analytically determined the ultimate sensitivities achievable
with the use of ancilla-assisted strategy and the optimal measurement
to saturate the sensitivities were also found. In contrast to the
ancilla-free strategy, we found that the effect of ancilla-assisted
strategy on sensitivity preserving is more pronounced for few number
of probes, but become negligible when the probe number is slightly
greater. Our results also indicate that taking a single ancillary
qubit in ancilla-assisted strategy is sufficient, while using more
ancillae will not bring any more benefit to frequency sensitivity.

\section*{Acknowledgements}
This work was supported by the NSFC through Grants No. 11747161, 11974189,
and a project funded by the Priority Academic Program Development
of Jiangsu Higher Education Institutions. L.Z. acknowledges support
from the China Postdoctoral Science Foundation under Grant No. 2018M642293.

\section*{Appendix A: Calculation of the QFIs \label{sec:AppA}}

  \makeatletter \renewcommand{\theequation}{A\arabic{equation}} \makeatother \setcounter{equation}{0}

In this appendix, we present detailed derivation of the QFI for ancilla-free
and ancilla-assisited frequency estimation strategies with the generalized
GHZ states in the presence of phase covariant noises. To facilitate
our calculation, we first obtain the analytical expression of the
QFI with respect to the phase of $\phi=\omega t$, denoted as $F_{\phi}$.
Then the QFI with respect to the frequency can be obtained from $F_{\phi}$
by using the equality of $F_{\omega}=t^{2}F_{\phi}$. In the scenarios
discussed in the main text, the phase-encoded density matrices can
be always decomposed into a direct sum of two sub-matrices, $\mathcal{E}\left(\rho_{{\rm GHZ}}\right)=\varrho_{\phi}\oplus\varrho_{D}$
with $\phi$-dependent $2\times2$ dimensional matrices $\varrho_{\phi}$
and $\phi$-independent diagonal matrices $\varrho_{D}$. According
to the sum property of the QFI \cite{Demkowicz-Dobrzanski2009PRA},
i.e., $F_{\phi}=\sum_{\nu}F\left(\varrho_{\nu}\right)$ for $\rho=\bigoplus_{\nu}\varrho_{\nu}$,
we find that the QFI of $\mathcal{E}\left(\rho_{{\rm GHZ}}\right)$
is simply given by $F_{\phi}=F\left(\varrho_{\phi}\right)$ due to
no contribution from the $\phi$-independent sub-matrix $\varrho_{D}$.
Thanks also to the simple form of $\varrho_{\phi}$, that is of dimension
only $2$, then its corresponding QFI can be formulated with respect
to the Bloch vector $\bm{r}$ of $\varrho_{\phi}$ as \cite{Zhong2013PRA}
\begin{eqnarray}
F\left(\varrho_{\phi}\right) & = & \frac{1}{r_{0}}\left[\vert\partial_{\phi}\bm{r}\vert^{2}+\frac{\left(\bm{r}\cdotp\partial_{\phi}\bm{r}\right)^{2}}{r_{0}-\vert\bm{r}\vert^{2}}\right],\label{eq:geo-QFI}
\end{eqnarray}
with the normalization factor  $r_{0}={\rm Tr\left(\varrho_{\phi}\right)}$.

At first, we consider the ancilla-free case in which the frequency
encoded density matrix is given by Eqs.~\eqref{eq:rho-2-free} and
\eqref{eq:rho-d-free}. By setting $\phi=\omega t$, we write the
$\phi$-dependent part as

\begin{eqnarray}
\varrho_{\phi} & = & \left(\begin{array}{cc}
2^{-N}\left(\left|c_{1}\right|^{2}A_{++}^{N}+\left|c_{2}\right|^{2}A_{-+}^{N}\right) & c_{1}c_{2}^{\ast}\eta_{\perp}^{N}e^{-iN\left(\theta+\phi\right)}\\
c_{1}^{\ast}c_{2}\eta_{\perp}^{N}e^{iN\left(\theta+\phi\right)} & 2^{-N}\left(\left|c_{1}\right|^{2}A_{--}^{N}+\left|c_{2}\right|^{2}A_{+-}^{N}\right)
\end{array}\right),\quad\quad
\end{eqnarray}
associating with
\begin{eqnarray}
r_{0} & = & 2^{-N}\left[\left|c_{1}\right|^{2}\left(A_{++}^{N}+A_{--}^{N}\right)+\left|c_{2}\right|^{2}\left(A_{-+}^{N}+A_{+-}^{N}\right)\right],\\
\bm{r} & = & \left(\begin{array}{c}
2c_{1}c_{2}^{\ast}\eta_{\perp}^{N}\cos N\left(\theta+\phi\right)\\
2c_{1}c_{2}^{\ast}\eta_{\perp}^{N}\sin N\left(\theta+\phi\right)\\
2^{-N}\left[\left|c_{1}\right|^{2}\left(A_{++}^{N}-A_{--}^{N}\right)+\left|c_{2}\right|^{2}\left(A_{-+}^{N}-A_{+-}^{N}\right)\right]
\end{array}\right).
\end{eqnarray}
By replacing Eq.~\eqref{eq:geo-QFI} with the above $r_{0}$ and
$\bm{r}$, one can directly obtain Eq.~\eqref{eq:QFI-noisyGHZ} as
a result of $\left(\bm{r}\cdotp\partial_{\phi}\bm{r}\right)^{2}=0$
and $F_{\omega}=t^{2}F_{\phi}$. Similarly, as for the ancilla-assisted
case, the $\phi$-dependent sub-density matrix corresponding to Eq.~\eqref{eq:rho-2-ancilla}
is
\begin{eqnarray}
\varrho_{\phi} & = & \left(\begin{array}{cc}
2^{-N}\left|c_{1}\right|^{2}A_{++}^{N} & c_{1}c_{2}^{\ast}\eta_{\perp}^{N}e^{-iN\left(\theta+\phi\right)}\\
c_{1}^{\ast}c_{2}\eta_{\perp}^{N}e^{iN\left(\theta+\phi\right)} & 2^{-N}\left|c_{2}\right|^{2}A_{+-}^{N}
\end{array}\right),
\end{eqnarray}
equipped with
\begin{eqnarray}
r_{0} & = & 2^{-N}\left(\left|c_{1}\right|^{2}A_{++}^{N}+\left|c_{2}\right|^{2}A_{+-}^{N}\right),\\
\bm{r} & = & \left(\begin{array}{c}
2c_{1}c_{2}^{\ast}\eta_{\perp}^{N}\cos N\left(\theta+\phi\right)\\
2c_{1}c_{2}^{\ast}\eta_{\perp}^{N}\sin N\left(\theta+\phi\right)\\
2^{-N}\left(\left|c_{1}\right|^{2}A_{++}^{N}-\left|c_{2}\right|^{2}A_{+-}^{N}\right)
\end{array}\right).
\end{eqnarray}
Submitting these into Eq.~\eqref{eq:geo-QFI}, one finally gets Eq.~\eqref{eq:QFI-ansillary-noisyGHZ}
as a consequence of $\left(\bm{r}\cdotp\partial_{\phi}\bm{r}\right)^{2}=0$
and $F_{\omega}=t^{2}F_{\phi}$.

\section*{Appendix B: Demonstration of the optimal observable \label{sec:AppB}}

  \makeatletter \renewcommand{\theequation}{B\arabic{equation}} \makeatother \setcounter{equation}{0}

While the QFI bounds the optimal sensitivity achievable with a given
quantum state, the sensitivity is limited by the measurements which
can be implemented in the experimental platform. Given a measurement
observable $O$, the attainable sensitivity is given by the error-propagation
formula
\begin{equation}
\Delta^{2}\hat{\omega}T\geq\frac{\left(\left\langle O^{2}\right\rangle -\left\langle O\right\rangle ^{2}\right)}{t\left|\partial\left\langle O\right\rangle /\partial\phi\right|^{2}},\label{eq:errorpro}
\end{equation}
where the sufficient and necessary condition for an observable $O$
to saturate the QFI bounds was proposed in \cite{Zhong2014JPA}.

Here, we show that the observable of direct-sum form $O_{A}=\sigma_{x}\oplus\bm{1}$
with $\sigma_{x}$ acting on the subspace of the density matrix corresponding
to $\varrho_{\phi}$ and $\bm{1}$ on the residual subspace corresponding
to $\varrho_{D}$, is an optimal measurement for both ancilla-free
and ancilla-assisted strategies discussed in the main text. Let us
take the ancilla-free strategy for example. The expectation values
of the first and second moments of $O$ in $\varrho_{\phi}\oplus\varrho_{D}$
can be easily calculated as
\begin{eqnarray}
\left\langle O\right\rangle  & = & 2\left|c_{1}\right|\left|c_{2}\right|\eta_{\perp}^{N}\sin N\left(\theta+\phi\right),\\
\left\langle O^{2}\right\rangle  & = & \left|c_{1}\right|^{2}\left(A_{++}^{2}+A_{--}^{2}\right)+\left|c_{2}\right|^{2}\left(A_{-+}^{2}+A_{+-}^{2}\right).
\end{eqnarray}
Submitting the above results into Eq.~\eqref{eq:errorpro} yields
\begin{equation}
\Delta^{2}\hat{\omega}T\geq\frac{\left(\left|c_{1}\right|^{2}\left(A_{++}^{2}+A_{--}^{2}\right)+\left|c_{2}\right|^{2}\left(A_{-+}^{2}+A_{+-}^{2}\right)-4\left|c_{1}\right|^{2}\left|c_{2}\right|^{2}\eta_{\perp}^{2N}\sin^{2}N\left(\theta+\phi\right)\right)}{4t\left|c_{1}\right|^{2}\left|c_{2}\right|^{2}\eta_{\perp}^{2N}N^{2}\cos N\left(\theta+\phi\right)}.
\end{equation}
Hence it meets the sensitivity bounds given by Eq.~\eqref{eq:QFI-noisyGHZ}
under the condition of $N\left(\theta+\phi\right)=2k\pi,k\in\mathbb{Z}$.

\end{document}